\documentclass[aps,prl,twocolumn,groupedaddress]{revtex4-1}  %<--------------------------------------------------
\usepackage{graphics}
\usepackage{epsfig}
\usepackage{color}

\begin{document}

%\preprint{}


\title{Carrier-wave Rabi flopping signatures in high-order harmonic generation for alkali atoms}

\author{M. F. Ciappina$^{1}$}
\email[]{marcelo.ciappina@mpq.mpg.de}
\author{J. A. P\'erez-Hern\'andez$^{2}$}
\author{A. S. Landsman$^{3}$}
\author{T. Zimmermann$^{4}$}
\author{M. Lewenstein$^{5,6}$}
\author{L. Roso$^{2}$}
\author{F. Krausz$^{1,7}$}

\affiliation{$^1$Max-Planck Institut f\"ur Quantenoptik, Hans-Kopfermann-Str.~1, D-85748 Garching, Germany}
\affiliation{$^2$Centro de L\'aseres Pulsados (CLPU), Parque Cient\'ifico, E-37008 Villamayor, Salamanca, Spain}
\affiliation{$^3$Max Planck Institute for the Physics of Complex Systems Nothnitzer Stra{\ss}e 38, D-01187 Dresden, Germany}
\affiliation{$^4$Physics Department, ETH Zurich, CH-8093 Zurich, Switzerland}
\affiliation{$^5$ICFO-Institut de Ci\`ences Fot\`oniques, Mediterranean Technology Park, 08860 Castelldefels (Barcelona), Spain}
\affiliation{$^6$ICREA-Instituci\'o Catalana de Recerca i Estudis Avan\c{c}ats, Lluis Companys 23, 08010 Barcelona, Spain}
\affiliation{$^7$Department f\"ur Physik, Ludwig-Maximilians-Universit\"at M\"unchen, Am Coulombwall 1, D-85748 Garching, Germany}

\date{\today}

\begin{abstract}
We present the first theoretical investigation of carrier-wave Rabi flopping in real atoms by employing numerical simulations of high-order harmonic generation (HHG) in alkali species.  Given the short HHG cutoff, related to the low saturation intensity, we concentrate on the features of the third harmonic of sodium (Na) and potassium (K) atoms. For pulse areas of 2$\pi$ and Na atoms, a characteristic unique peak appears, which, after analyzing the ground state population, we correlate with the conventional Rabi flopping. On the other hand, for larger pulse areas, carrier-wave Rabi flopping occurs, and is associated with a more complex structure in the third harmonic.  These new characteristics observed in K atoms indicate the breakdown of the area theorem, as was already demonstrated under similar circumstances in narrow band gap semiconductors.


\end{abstract}

%\pacs[PACS numbers]{42.65.Ky, 78.67.Bf, 32.80.Rm}
%\item{PACS numbers: 42.65.Ky, 78.67.Bf, 32.80.Rm}
%\pacs{42.65.Ky, 78.67.Bf, 32.80.Rm}
%\keywords{}
%\subsection{PACS codes}
%If supplying suggested PACS codes, they must be supplied as a
%comma-separated list using a single \m{pacs} macro.
\maketitle

It is well known that semiconductors, when modeled as a two-level system, develop a periodic oscillation of the population inversion when interacting with constant light, a phenomenon predicted by I. I. Rabi in the 30s, called Rabi flopping~\cite{rabi}. Rabi flopping has also been observed when using ultrafast optical pulses e.g.~\cite{cundiff,hughes}. For these pulses, peculiar behavior emerges when the driven light intensity is so high that the period of one Rabi oscillation is comparable with that of one cycle of light.  In this case, the area theorem has been shown to break down~\cite{hughes}, and a new phenomena, known as carrier-wave Rabi flopping (CWRF), emerges. These features can be schematically observed in the so-called Bloch sphere (for the definition and more details see e.g.~\cite{wegener}), presented in Fig.~1. 

In particular, Fig.~1(a) depicts the \textit{conventional} Rabi flopping on a Bloch sphere. For this case the Rabi period is much larger than the driven light period and the Bloch vector, formed by the real ($u$) and imaginary ($v$) parts of the optical polarization and the population inversion ($w$) of a two-level system, spirals up starting from the south pole (corresponding to all the electrons in the ground state), reaches the north pole and returns to its initial position for the case of squared-shaped pulses with an envelope area of $\Theta=2\pi$. Here optical oscillations are mapped to an orbit of the Bloch vector parallel to the $uv$ or equatorial plane. Additionally, oscillations of the population inversion are given by the motion in the $uw$ plane. The corresponding spectrum of the optical polarization would exhibit then two peaks centered around the two-level transition frequency. On the other hand, Fig.~1(b) presents results for a much shorter pulse, where the Rabi period is equal to the driven light period. Even when the envelope area for this case is $\Theta=4\pi$, it is clear that the Bloch vector does not return to the south pole, as may be expected.  To the contrary, a more chaotic behavior is observed in the motion of the Bloch vector, resulting in a more complex shape in the spectrum of the optical polarization. Furthermore, the well-known area theorem of the nonlinear optics fails when this parameter regime is reached. Note that multipeak splitting of the resonance fluorescence spectrum by short pulses in the standard Rabi flopping regime was predicted in Refs.~\cite{maciej1,maciej2,maciej3,maciej4}, although this effect is due to a complex temporal interference effect, rather than CWRF.

\begin{figure}[ht]
\resizebox{2.4in}{!}{\includegraphics[angle=0]{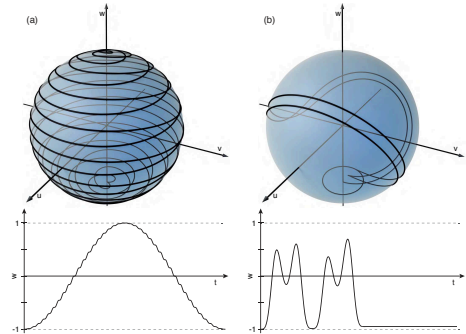}}  
\caption{Sketch of the Bloch Sphere showing the different regimes. (a) schematic showing the \textit{travel} of the Bloch vector for conventional Rabi flopping for a pulse with an envelope pulse are $\Theta=2\pi$. (b) same for carrier-wave Rabi flopping for a pulse with an envelope pulse area of $\Theta=4\pi$. The bottom panels show the evolution of the population inversion $w$ (see the text for more details). }
\label{fig1}
\end{figure}

\newpage

\begin{figure*}[t]
\resizebox{4.5in}{!}{\includegraphics[angle=0]{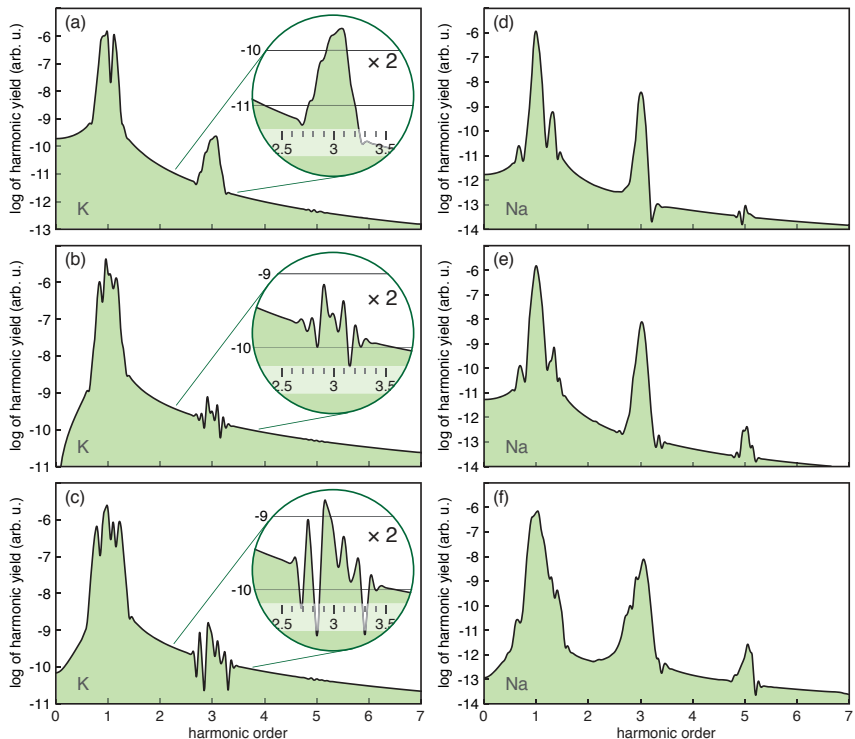}}  
\caption{3D-TDSE harmonic spectra in K for the corresponding laser intensities $I=3.158\times10^{11}$ W/cm$^{2}$ (panel a), $I=5.6144\times10^{11}$ W/cm$^{2}$ (panel b) and $I=1.108\times10^{12}$ W/cm$^{2}$ (panel c). Panels (d), (e) and (f) represent the HHG in Na for the same laser parameters. The insets of panels (a), (b) and (c) show a zoom of the third harmonic  $\omega/\omega_0=3$ (see the text for details).}
\label{fig2}
\end{figure*}

Experiments on narrow band gap semiconductors have shown a clear signature of CWRF, which manifested itself as a split in the third harmonic of the emitted light into the forward direction~\cite{muckeexp}. Recently, Rabi flopping and the consequent coherent pulse reshaping has been experimentally observed in a quantum cascade laser~\cite{choi}, suggesting a new promising approach to short pulse generation. One of the advantages of atoms (relative to semiconductors) is the possibility to employ longer laser pulses and, as a consequence, to explore a broader range of laser parameters, as well as provide an alternative to carrier envelope phase (CEP) characterization.  In addition, it has been shown that in semiconductors the Coulomb interaction of carriers in the bands gives rise to an enhancement of the external laser field and consequently the envelope pulse area by as much as a factor of two, considerably complicating the observed interpretation of CWRF phenomena \cite{muckeexp,cundiff,ciuti}.  

When the atom was simplistically modeled as a two-level system, conventional Rabi flopping behavior and CWRF features were observed (see e.g.~\cite{wegener,frasca} and references therein). However, it is well known that the two-level approximation breaks down when strong electric fields are applied, in particular in the CWRF regime.
An important question emerges as to what extend CWRF could potentially be observed in real atoms. In this Letter we demonstrate for the first time how the CRWF signatures show up in the high-order harmonic generation (HHG) spectra of real atoms.  In particular, using a robust theoretical approach that accurately models {\it both the ground and excited states} of K atoms combined with realistic laser parameters, we observe clearly distinct features in the third harmonic and correlate it with the behavior of the ground state population.  
The latter approach is closely related to the description of semiconductors and atoms modeled as two-level systems.  In order to further support our conclusions of CWRF-like behavior in K, we also compute HHG of Na atoms, in which the transition energy between the ground and first excited state is not resonant with the driven light, and, as a consequence, a conventional HHG spectra, i.e. single peaks at odd harmonics of the driven frequency, is observed.
Our predictions can be tested experimentally using currently available ultrashort laser pulses of a Ti:Sa laser with wavelengths centered in the range of $750-800$ nm.

We start by describing our theoretical approach, putting special emphasis on the choice of the atomic potentials, such that our results for both ground and excited states are in excellent agreement with experimental measurements (see Table I). To find CWRF signatures, we focus on the HHG spectra. Since the HHG spectra is proportional to the electron dipole moment, we can establish a one to one correspondence between the media, modeled as a collection of oscillators, and single atoms illuminated by a strong laser field (any macroscopic effect, such as phase matching, could be safely neglected, considering the low-order harmonic cutoff developed in alkali atoms, closely related to their low saturation intensity). To create the conditions for CWRF, we used an atomic system in which the period of a Rabi oscillation~\cite{eberly} (corresponding to the transition between the ground and the first excited states ~\cite{milonni,meystre}) is similar to one period of the laser light. For the usual (Ti:Sa) laser sources such system was provided by K atoms, with a transition energy between the ground and the first excited state of 1.61 eV (hence close to the laser source photon energy of 1.55 eV).

Alkali atoms, due to their atomic structure (gas noble structure plus only one \textit{external} electron), are well-suited to be described by the single active electron approximation (SAE). We therefore focus on such atoms to avoid the possible role of electron-electron correlations, which have been found to have an important, yet still poorly understood role in HHG spectra~\cite{shiner}. Based on SAE approximation, we use the atomic potential reported in~\cite{perez-hernandez04} to describe K and Na atoms. Using a Hartree-Fock-based method we set the two parameters $V_c$ and $V_e$ in the following generic potential form, $V_{K,Na}(r)={V_c\over(r+r_0)^2}-{V_e\over (r+r_0)}$, where $V_c$ accounts for the effect of the atomic core (nucleus plus all complete shells), $V_ e$ represents the external potential and $r_0={V_c\over V_e}$.  Using this method we find the ground, $3s$, and the first excited state, $3p$, of K, as well as the $4s$ and $4p$ for Na, with  precision of $\Delta E\approx\pm0.0084$eV (for details see~Table I). In addition, we also compute numerically the element transition dipole $d_{ns\rightarrow np}=\langle \psi_{ns} | z | \psi_{np}\rangle$ for both atoms ($n=3$ for Na and $n=4$ for K) and compare them with the experimental values reported in Ref.~\cite{steck} (see~Table I). As can be deduced, our theoretical dipole values show excellent agreement with experimental measurements, indicating that the potential given above accurately represents K and Na atoms. For finding optimal laser parameters to experimentally observe CWRF in atoms, an accurate dipole moment is essential since Rabi flopping frequency is linearly proportional to the dipole moment as well as the electric field strength~\cite{eberly,choi}. In addition, HHG spectra, which is the focus of our investigation, is believed to be particularly sensitive to the details of electron dynamics inside atoms and molecules, making it crucial to not only accurately describe the ground state (as many atomic potentials in the literature do), but, in the case of resonant transitions due to Rabi flopping, also the excited state~\cite{shiner,boguslavskiy}.

\begin{table}
%\begin{ruledtabular}
\begin{tabular}{p{1.20in}p{1.20in}p{0.90in}} \hline
Sodium  & Experimental (NIST) & Numerical (present)  \\ \hline
$3s$  & 5.139 eV & 5.135 eV  \\
$3p$ & 3.036 eV & 3.038 eV \\
Transition dipole & 2.49 au & 2.40 au \\ \hline
%\end{tabular}
%\label{table:tdse_scalings}
%\end{table}

%\begin{tabular}{p{1.20in}p{1.20in}p{0.90in}}\hline
Potassium  & Experimental (NIST) & Numerical (present) \\ \hline
$4s$  & 4.340 eV & 4.347 eV  \\
$4p$ & 2.730 eV & 2.725 eV \\
Transition dipole & 2.92 au & 2.79 au \\ \hline
%\end{ruledtabular}
\end{tabular}
\label{table:table}
\caption{Experimental and theoretical values of the energy gap between the ground and first excited states, joint with the transition dipole both for Na and K.}
\end{table}

%Table~\ref{tab:fonts}.
%
%\begin{table}
%\caption{\label{tab:fonts} Sodium}
%\begin{ruledtabular}
%\begin{tabular}{lp{2in}}
%3s & 5.139 & 5.135 eV
%3p & 3.036 eV & 3.038 eV
%dipole amplitude
%\end{tabular}
%\end{ruledtabular}
%\end{table}
%

To compute HHG spectra, we numerically solve the three dimensional Time Dependent Schr\"odinger Equation (3D-TDSE) in the length gauge and using the atomic potential, $V_{K,Na}(r)$, given above for K and Na atoms, respectively. The harmonic yield from a single atom is then proportional to the Fourier transform of the dipole acceleration of its active electron and can then be obtained from the electronic wave function after time propagation. 
Our code is based on an expansion of spherical harmonics, $Y_{l}^{m}$, and takes advantage of the cylindrical symmetry of the problem (hence only the $m = 0$ terms need to be considered). The time propagation is based on a Crank-Nicolson method implemented on a splitting of the time-evolution operator that preserves the norm of the electronic wave function. The coupling between the atom and the laser pulse in the length gauge, linearly polarized along the $z$ axis, is written as $V_{l}(z,t)=E(t)\,z$, where $E(t)$ is the laser electric field defined by $E(t)=E_0 \sin^2\left( { \omega_0 t \over 2N }\right) \sin (\omega_0 t+\phi)$. $E_0$ is the laser electric field peak amplitude ($E_0=\sqrt{I/I_0}$ with $I_0=3.5\times 10^{16}$ W/cm$^2$), $\omega_0=0.0596$ a.u. ($\lambda=765.1$ nm), $N$ the total number of cycles in the pulse and $\phi$ the CEP. Furthermore, $T$ defines the laser period $T=2\pi/\omega_0\approx 2.5$ fs. In the simulations presented here we consider the case $N = 20$, corresponding to an intensity envelope of full width half maximum (FWHM) of $0.36 N T$, (7.2 optical cycles$\approx 18$ fs FWHM), and $\phi=0$ (see below for details). 

As discussed in the introduction, we set input laser frequency the same value (in atomic units) as the energy corresponding to the transition $4s\rightarrow4p$ of K in order to observe a CWRF-like behavior. In order to compare with a {\it conventional} situation (meaning the usual conditions for HHG), we use the same input laser parameters with Na atoms, for which the transition energy from the ground to the first excited state, $3s\rightarrow3p$, corresponds to 2.10 eV, and is therefore non-resonant with the laser frequency.

\begin{figure*}[ht]
\resizebox{5in}{!}{\includegraphics[angle=0]{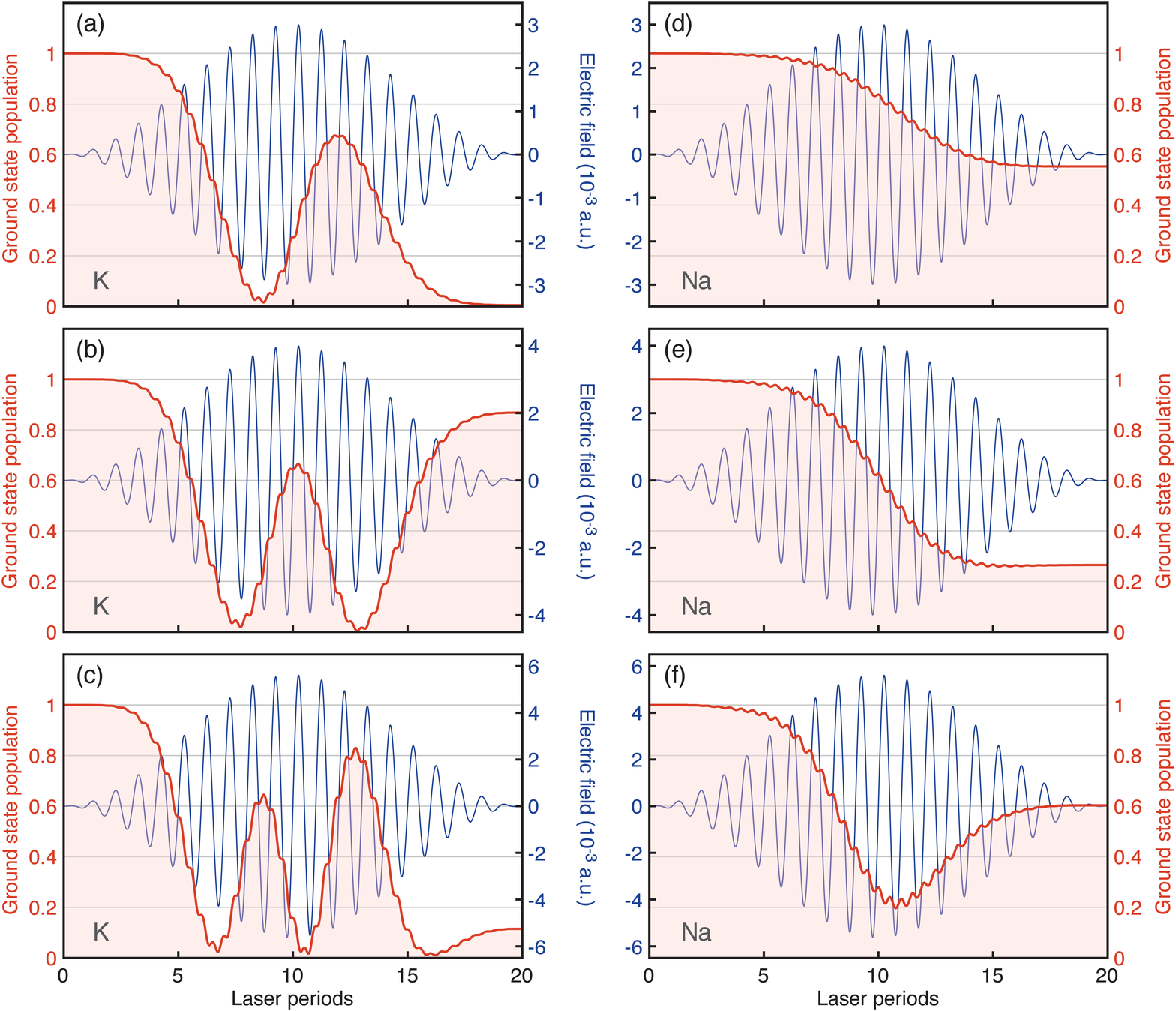}}  
\caption{Time evolution of the ground state population (red thick line) along the laser pulse (blue thin line) corresponding to the cases plotted in Fig.~2.}
\label{fig3}
\end{figure*}

In Fig.~2, we show the harmonic spectra computed from the 3D-TDSE for both K (Figs.~2(a), 2(b) and 2(c)) and Na (Figs.~2(d), 2(e) and 2(f)) atoms. We have chosen the laser parameters to cover three different regimes, namely, for panels 2(a) and 2(d) the envelope pulse area, $\Theta_{K,Na}$, is close to $2\pi$.  The envelope pulse area is defined as $\Theta_{K,Na}\approx d_{K,Na} E_0 \Delta t$, where $d_{K,Na}$ is the dipole transition matrix for K or Na (see Table I) and $\Delta t$ the FWHM pulse duration ($\Delta t=18$ fs i.e., $\Delta t\approx 750$ au). In particular, for a laser intensity $I=3.158\times10^{11}$ W/cm$^{2}$ ($E_0=0.003$ a.u.), $\Theta_K\approx 2\pi $ and $\Theta_{Na}\approx 5.4$ (for comparison see the values used in semiconductor GaAs~\cite{muckeexp}). 
Panels 2(c) and 2(f) correspond to values of $\Theta_K$ and $\Theta_{Na}$, respectively, close to $4\pi$ (11.7 for K and 10.1 for Na), obtained by keeping constant the pulse duration and now using a laser intensity $I=1.108\times10^{12}$ W/cm$^{2}$ ($E_0=0.0056$ a.u.). 
Panels 2(b) and 2(e) were chosen to have an intermediate value of intensity $I=5.6144\times10^{11}$ W/cm$^{2}$ ($E_0=0.004$ a.u.), corresponding to pulse envelope areas of 8.4 and 7.2 for K and Na, respectively. 

From the HHG spectra of K atoms, we observe a drastic change around the third harmonic, $\omega/\omega_0 =3$, as pulse envelope area increases.  Our results are directly analogous to the manifestation of CWRF behavior observed in semiconductors in~\cite{muckeexp}. Note the marked contrast to the HHG spectra of Na, shown in panels 2(d), 2(e), and 2(f), where a characteristic peak is present in the third harmonic regardless of the envelope pulse area. The onset of this more complex behavior in K atoms for sufficiently large envelope pulse areas is in agreement with conclusions in \cite{hughes}, where the onset of CWRF behavior and the consequent break-down of the area theorem was predicted for a two-level resonant systems when the pulse envelope area significantly exceeds $2\pi$.

To get further insight into the physical mechanism behind the complex structure of the HHG spectra in K atoms, in Fig.~3 we present the time dynamics of the ground state population for all the cases depicted in Fig.~2.  
For a two-level system, used as a prototypical model for a semiconductor or a simplistic picture for a real atom, the electron dynamics due to interaction with laser light can be represented on a Bloch sphere (for details see e.g~\cite{wegener,muckeexp}).  In this case, the ground state populations and the regular Rabi oscillations can be depicted as moving along the surface of the sphere (see Fig.~1(a)).
When the CWRF regime is reached, clear signatures, corresponding to the break-down of the area theorem, should occur on the Bloch sphere as well (see Fig.~1(b)).
  
Following an analogy with a two-level system, we can distinctly observe CWRF-like behavior in Figs.~3(b) and 3(c) and the corresponding counterpart in the HHG spectra (Figs.~2(b) and 2(c)). On the contrary, a \textit{conventional} behavior in the third harmonic (Figs.~2(a) and 3(a)), can be correlated with: (i) ordinary Rabi oscillations for the case of K (Fig.~2(a)), i.e. the ground state is completely depopulated, even though the laser intensity is low and this would be analogous to a \textit{travel} of the Bloch vector from the south to the north pole~\cite{muckeexp}; (ii) normal behaviour of atoms in strong field for all the Na cases (Figs.~2(d)-2(f)), i.e. gradual depopulation of the ground state due to laser ionization (Figs.~3(d)-3(f)).  

\begin{figure}[h]
\resizebox{3in}{!}{\includegraphics[angle=0]{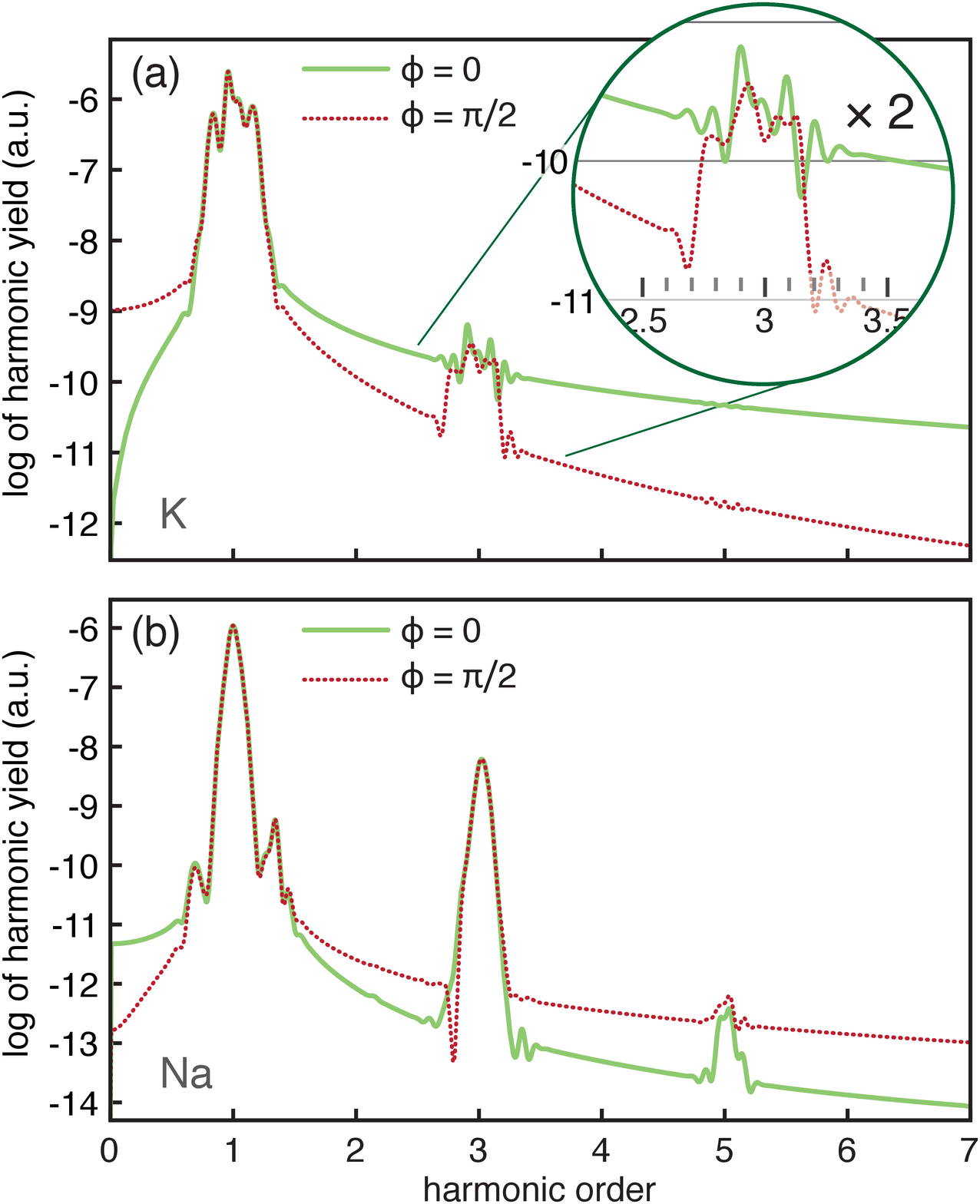}}  
\caption{HHG for K (panel a) and Na (panel b) for different CEPs. The laser parameters are the same as in Figs.~2(b) and 2(e), respectively. Solid line $\phi=0$, dotted line $\phi=\pi/2$.}
\label{fig4}
\end{figure}

In conclusion, we find the signatures of CWRF in real atoms, by studying the third harmonic of alkali atomic species. Analogous to the case of semiconductors, we can correlate this new feature with the complex dynamics of the ground state population. Our model uses accurate values for the atomic wavefunction of both ground and excited states (as is evidenced by the excellent agreement between the calculated states energies with experimental values) as well as accurate laser parameters, easily achievable with the current laser technology. In particular, a Ti:Sa laser provides laser pulses with wavelengths centered in the range $750-800$ nm, very close to the $765$ nm value corresponding to the transition energy ${4s\rightarrow 4p}$ in K. As a consequence, the experimental confirmation of our results appears straightforward. Moreover, the CWRF phenomenon in atoms could emerge as a robust alternative for CEP characterization for long pulses as can be seen in Fig.~4 where we show HHG spectra for K (4(a)) and Na (4(b)) and for two different values of the CEP. Note that in the case of Na (out of resonance) the third harmonic does not present appreciable differences. However the third harmonic of K is strongly affected, in spite of the fact the driving laser is rather long (20 cycles) of total duration. It is well known that in a {\it conventional} situation the HHG spectra is only sensitive to the CEP changes when the driving laser field is a few-cycle pulse. Furthermore the CWRF could be used to control the laser-induced ionization, known as a crucial ingredient for harmonic propagation, via manipulation of the ground state population.   

\begin{acknowledgments}
We acknowledge the financial support of the MICINN projects (FIS2008-00784 TOQATA, FIS2008-06368-C02-01, and FIS2010-12834), ERC Advanced Grant QUAGATUA and OSYRIS, the Alexander von Humboldt Foundation (M.L.), and the DFG Cluster of Excellence Munich Center for Advanced Photonics. This research has been partially supported by Fundaci\`o Privada Cellex. J.A.P.-H. and L. Roso acknowledge support from Laserlab-Europe (Grant No.~EU FP7 284464) and the Spanish Ministerio de Econom\'ia y Competitividad (FURIAM Project FIS2013-47741-R). We thanks Christian Hackenberger for helping us with the artwork.
\end{acknowledgments}



%\label{sec:TeXbooks}

%\newpage

%Figures Captions

%Fig.~1: Sketch of the Bloch Sphere showing the different regimes. (a) schematic showing the \textit{travel} of the Bloch vector for conventional Rabi flopping. (b) same for carrier-wave Rabi flopping (see text for more details). 

%Fig.~2: 3D-TDSE harmonic spectra in potassium (K) for the corresponding laser intensities $I=3.158\times10^{11}$ W/cm$^{2}$ (panel a), $I=5.6144\times10^{11}$ W/cm$^{2}$ (panel b) and $I=1.108\times10^{12}$ W/cm$^{2}$ (panel c). Panels (d), (e) and (f) represent the HHG in sodium (Na) for the same laser parameters. The insets of panels (a), (b) and (c) show a zoom of the third harmonic  $\omega/\omega_0=3$ (see the text for details).

%Fig.~3: Time evolution of the ground state population (red thick line) along the laser pulse (blue thin line) corresponding to the cases plotted in Fig.~2.

\end{document}